\documentclass[sigconf,nonacm]{acmart}

\usepackage{booktabs} % For formal tables

\usepackage{listings}
\usepackage{xcolor}

\definecolor{codegreen}{rgb}{0,0.6,0}
\definecolor{codegray}{rgb}{0.5,0.5,0.5}
\definecolor{codepurple}{rgb}{0.58,0,0.82}
\definecolor{backcolour}{rgb}{0.95,0.95,0.92}

\lstdefinestyle{mystyle}{
    backgroundcolor=\color{backcolour},   
    commentstyle=\color{codegreen},
    keywordstyle=\color{magenta},
    numberstyle=\tiny\color{codegray},
    stringstyle=\color{codepurple},
    basicstyle=\ttfamily\tiny,
    breakatwhitespace=false,         
    breaklines=true,                 
    captionpos=b,                    
    keepspaces=true,                 
    numbers=left,                    
    numbersep=5pt,                  
    showspaces=false,                
    showstringspaces=false,
    showtabs=false,                  
    tabsize=2
}

\usepackage[ruled]{algorithm2e} % For algorithms

\SetAlFnt{\small}
\SetAlCapFnt{\small}
\SetAlCapNameFnt{\small}
\SetAlCapHSkip{0pt}

\begin{document}

\title[Non-linear, Team-based VR Training for Cardiac Arrest Care with enhanced CRM Toolkit]
{Non-linear, Team-based VR Training for Cardiac Arrest Care with enhanced CRM Toolkit}

\begin{teaserfigure}
  \mbox{}
  \hfill
  \includegraphics[width=0.32\textwidth, height=100pt]{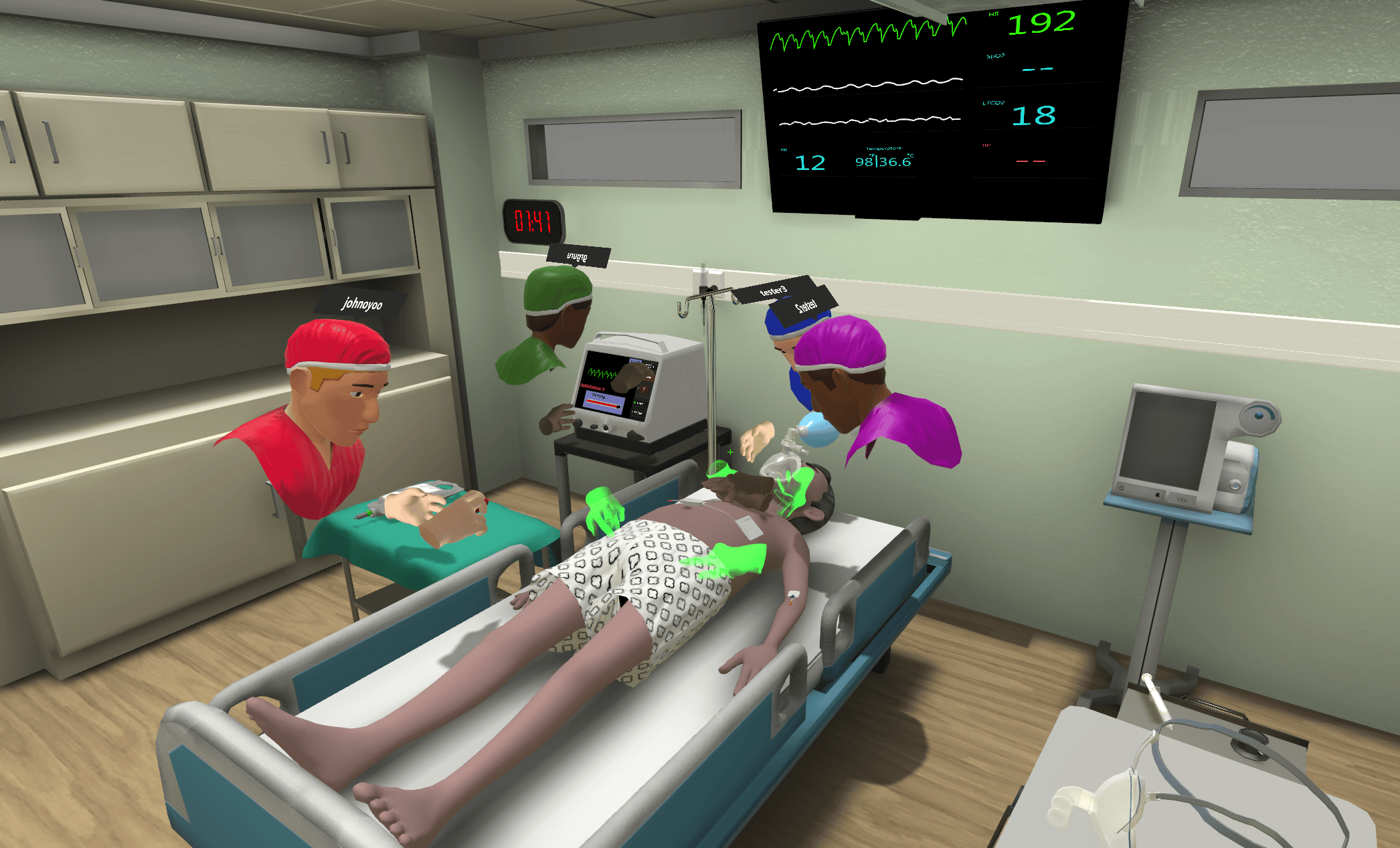}
  \hfill
  \includegraphics[width=0.32\textwidth, height=100pt]{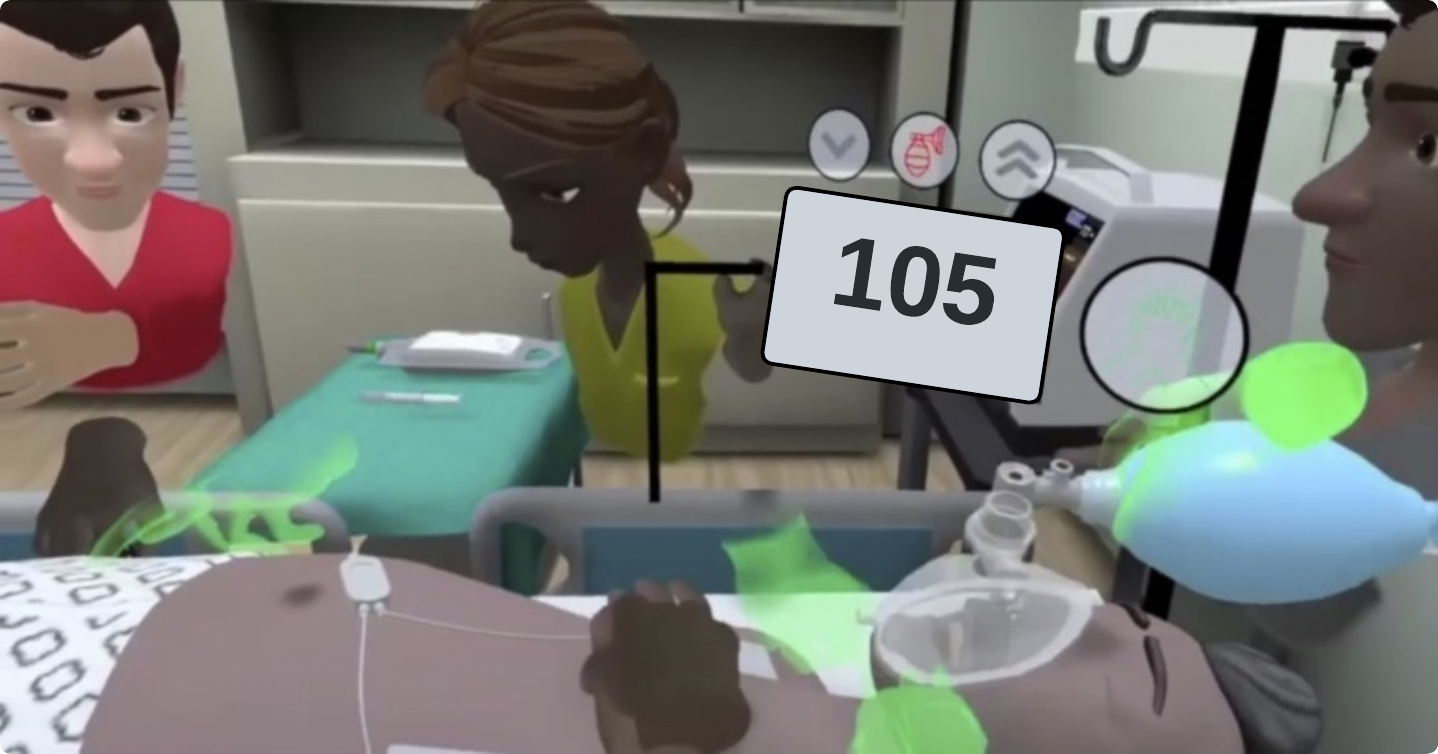}
  \hfill 
  \includegraphics[width=0.32\textwidth, height=100pt]{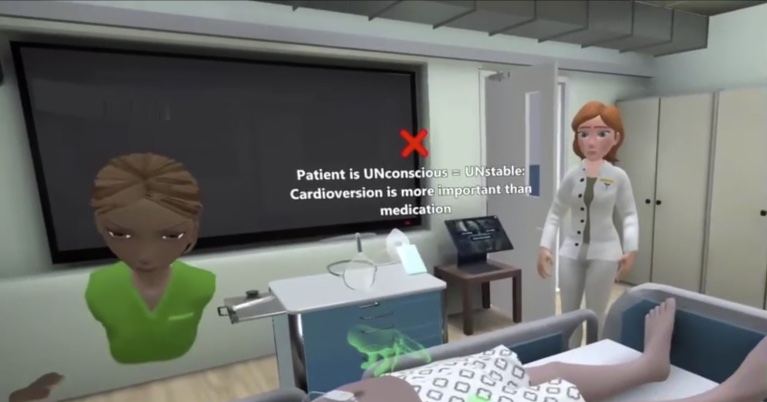}
  \hfill
  \mbox{}
  \centering
  \caption{%
         (Left) A collaborative VR session where trainees treat a cardiac arrest patient, performing actions based on care protocols. The patient’s condition changes in real-time, reacting to performed actions with vital signs feedback. (Middle) A trainee adjusts CPR compression rate while another uses a bag valve mask. (Right) Trainees receive non-overwhelming modulated real-time feedback on clinical errors.
         }
    \Description{}
  \label{fig:teas1}
\end{teaserfigure}

\author{Mike Kentros}
\orcid{0000-0002-3461-1657}
\affiliation{%
  \institution{FORTH - ICS, University of Crete, ORamaVR}
  \city{}
  \country{}
}
\author{Manos Kamarianakis}
\orcid{0000-0001-6577-0354}
\affiliation{%
  \institution{FORTH - ICS, University of Crete, ORamaVR}
  \city{}
  \country{}
}
\author{Michael Cole}
\orcid{0009-0000-4146-3632} 
\affiliation{%
  \institution{University of Michigan Medical School}
  \city{}
  \country{}
}

\author{Vitaliy Popov}
\orcid{0000-0003-2348-5285}
\affiliation{%
  \institution{University of Michigan Medical School}
  \city{}
  \country{}
}

\author{Antonis Protopsaltis}
\orcid{0000-0002-5670-1151}
\affiliation{%
  \institution{University of Western Macedonia, ORamaVR}
  \city{}
  \country{}
}
\author{George Papagiannakis}
\orcid{0000-0002-2977-9850}
\affiliation{%
  \institution{FORTH - ICS, University of Crete, ORamaVR}
  \city{}
  \country{}
}
\renewcommand{\shortauthors}{Kentros, Kamarianakis et al.}

\begin{abstract}

This paper introduces iREACT, a novel VR simulation addressing key limitations in traditional cardiac arrest (CA) training. Conventional methods struggle to replicate the dynamic nature of real CA events, hindering Crew Resource Management (CRM) skill development. iREACT provides a non-linear, collaborative environment where teams respond to changing patient states, mirroring real CA complexities. By capturing multi-modal data (user actions, cognitive load, visual gaze) and offering real-time and post-session feedback, iREACT enhances CRM assessment beyond traditional methods. A formative evaluation with medical experts underscores its usability and educational value, with potential applications in other high-stakes training scenarios to improve teamwork, communication, and decision-making.

\end{abstract}



\maketitle

\section{Introduction}

Hospital-based cardiac arrest (CA) is a time-critical medical emergency procedure, requiring rapid and coordinated intervention by a multi-disciplinary medical team \cite{Perkins2015European}. Current CA training methods, often relying on static mannequins and didactic lectures, struggle to replicate the dynamic and unpredictable nature of real-world CA events \cite{Cheng2018Resuscitation}. First, they often follow linear, pre-scripted scenarios, failing to adequately prepare trainees for the unexpected events and rapidly changing patient states characteristic of actual CA situations, hindering the development of crucial adaptive decision-making skills. Second, current manual assessment methods for 
(CRM) skills, such as direct observation by a single instructor, are prone to human error, difficult to standardize, and challenging to replicate \cite{Cheng2018Resuscitation}. Observing and logging actions of multiple trainees in a fast-paced environment is cognitively demanding for instructors, leading to potential oversights and inconsistencies in evaluation. Since these methods often lack objective, quantifiable data for team-based tasks, they struggle to effectively assess team performance. 
These training methods are hard to scale due to the need for physically present instructors.

To address these challenges, we designed iREACT (Immersive virtual Reality Environment for training Acute Care Teams), an innovative VR simulation, developed with medical experts, that enhances CA training by overcoming traditional methods' limitations. iREACT offers a non-linear, collaborative environment where four trainees respond to dynamically changing, clinically accurate patient states, mirroring real-world CA complexities. This setup forces trainees to adapt and make real-time decisions. iREACT also introduces a new approach to CRM assessment by capturing multi-modal data (user actions, cognitive load, visual gaze) and providing real-time and post-session feedback, eliminating the limitations of traditional methods.
iREACT introduces an advanced assessment approach, integrating the well-documented traditional CA training assessment of communication, behavior, and clinical decision-making \cite{macnamara2021high,beal2017effectiveness} into a collaborative VR environment. It combines learning data from multiple modalities at both individual and team levels into a unified training system. The principles and methodology \cite{MAGES4} used in iREACT can be generalized to a broader range of team-based scenarios, enhancing training in other high-stakes domains.

% %%%%%%%%%%%%%%%%%%%%%%%%%%%%%%%%%%%
\section{iREACT - Designed by Medical Experts} % (fold)
\label{sec:iREACT_Overview}

iREACT was designed through an extensive, iterative process in close collaboration with a team of medical experts to address the specific needs and challenges of CA training. This collaboration spanned several months and involved continuous feedback in the spiral system-design process. 
The expert team, including doctors from the University of Michigan Medical School, guided design and development to ensure clinical accuracy, effective teaching, and ACLS alignment. Their feedback on early VR prototypes shaped scenarios, UI, and interactions to meet training goals.

This process resulted a \textit{pure} multi-user collaborative gameplay, allowing a team of exactly four VR trainees, each one assigned a unique medical role, to cooperate and provide essential and effective handling of a patient in CA (see Figure~\ref{fig:teas1}). For this reason, iREACT involves a sandbox gameplay environment, providing a variety of decision options for the trainee with a set of more than 20 available medical actions, where trainees can decide and choose the most appropriate one to treat the patient’s dynamically changing clinical condition (e.g., perform chest compressions, administer medications, use medical tools such as stethoscopes, EKGs, and X-rays).
Adaptive game dynamics with patient vital signs changing in response to the trainees’ actions, reflecting real-world patient conditions, allow trainees to monitor and respond to the evolving condition of the patient and understand the impact of their actions. The combination of these features mimics the collaborative and non-linear nature of real-life medical emergencies and assist trainees in developing their decision-making abilities. Moreover, as iREACT integrates cognitive load and eye tracking functionalities it monitors the trainee's learning impact and provides user-specific instructional feedback.
To the authors' knowledge, iREACT is the first non-linear, team-based VR simulation for immersive  and engaging healthcare training.

An innovative set of meticulously crafted learning tools, handpicked by medical experts, advances the training experience even more. Enabled by the MAGES SDK \cite{MAGES4}, these features include (i) real-time feedback during the training simulation, (ii) robust analytics, (iii) guided review in-VR after the training simulation, (iv) session video recording/replaying able to be viewed from various perspectives (see Figure~\ref{fig:replays}-Bottom) to facilitate engaging, instructor-led post-session debriefing 
, (v) quantitative assessment of team communication and coordination, and (vi) cognitive load and eye gaze tracking (described below).

\section{Features and Design Choices} 
\label{sec:Features_and_Design_Choices}

Evidence demonstrates the impact of cognitive and behavioral factors on a team’s ability to provide effective CA care \cite{Chan2016}. iREACT leverages the unique capabilities of VR to address these factors through several key features and design choices:
\begin{figure}[tbp]
    \centering
    \includegraphics[width = 0.95 \linewidth]{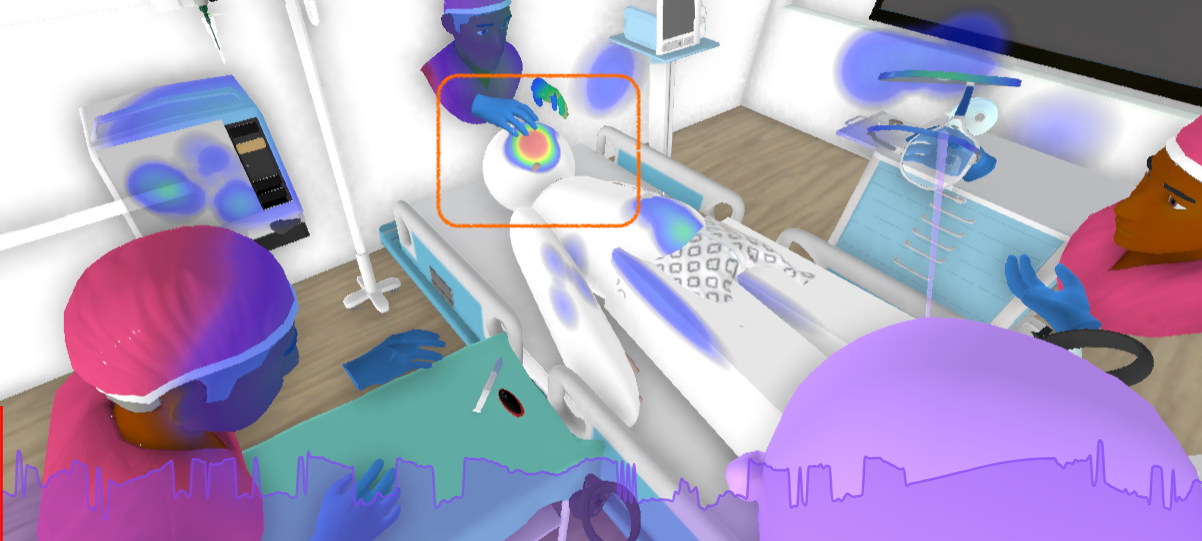}
    \includegraphics[width = 0.95 \linewidth]{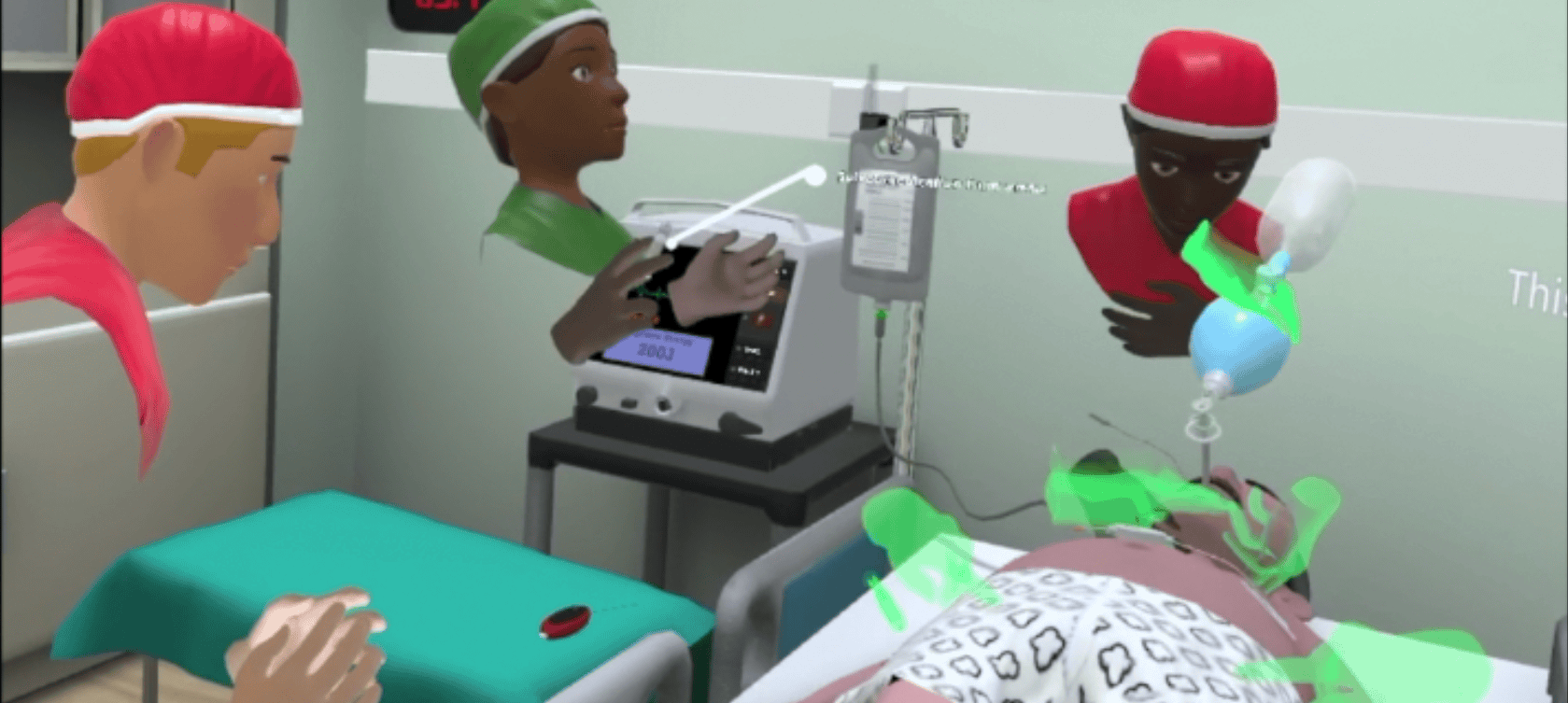}
    \caption{
(Top) Cognitive load (purple) and gaze heatmap (orange) help identify errors.
(Bottom) Replay from a new angle aids trainee and instructor review to prevent future mistakes.}
    \Description{}
    \label{fig:replays}
\end{figure}

\textbf{Non-Linear Scenario Design:} A core feature of iREACT is its non-linear scenario design. Unlike traditional training methods that follow pre-scripted sequences, iREACT presents dynamic patient states that change in response to trainee actions. The patient’s clinical condition is governed by a state machine that transitions between different states (e.g., ventricular fibrillation, asystole, return of spontaneous circulation) based on both pre-programmed events and trainee interventions. 
Trainee actions affect vitals and trigger state changes, creating a branching narrative that builds adaptive decision-making in high-pressure scenarios.

\textbf{An exclusively Multi-User Simulation:} 
 iREACT is a strictly multi-user simulation requiring four trainees with defined medical roles: team leader, chest compressions, airway management, and defibrillator/medication administration. This pure collaborative setup ensures optimal user synchronization through the authoring framework’s network capabilities \cite{MAGES4}. The solution has been tested and shown to function effectively in co-op sessions, using decoupled cloud-based physics server over 5G networks \cite{protopsaltis2024}.
 While unlimited viewers can observe, the participation of four active trainees is a medical training requirement. Real scenarios with fewer responders involve multitasking. 
 Adapting the simulation for flexible participant numbers requires major design changes and is planned for future work.

\textbf{Multi-Modal Data Capture and Integration:} iREACT integrates data from multiple sources to provide comprehensive feedback on trainee performance. This includes:
\begin{itemize}
    \item \textbf{User Actions:} All actions performed by trainees within the VR environment are recorded with precise timestamps.
    \item \textbf{User Gaze and Hearth Rate} Using the heart rate, eye tracking and pupillometry sensor of a specific VR HMD (HP Reverb G2 Omnicept Edition), 
    all user data are recorded in real-time. 
    \item \textbf{Cognitive Load:} 
         User cognitive load is estimated on the HP HMD from recorded data using the HP Omnicept SDK (\url{https://developers.hp.com/omnicept}), by utilizing user's eye gaze, heart rate, pupillometry, and temperature. By utilizing Cognitive3D SDK (\url{www.cognitive3d.com}), we provide web-based replay of the session with visualization of cognitive load and gaze behavior (see Figure~\ref{fig:replays}-Top). Future work aims to develop a more adaptive, transparent cognitive load evaluation tool.
    \item \textbf{Speech Data:} 
    Trainee speech is recorded and transcribed to reveal team dynamics and communication effectiveness.
\end{itemize}
This multi-modal data is time-synchronized and stored for post-session analysis and replay. 

\begin{figure*}
    \centering
    \includegraphics[width=0.95\textwidth]{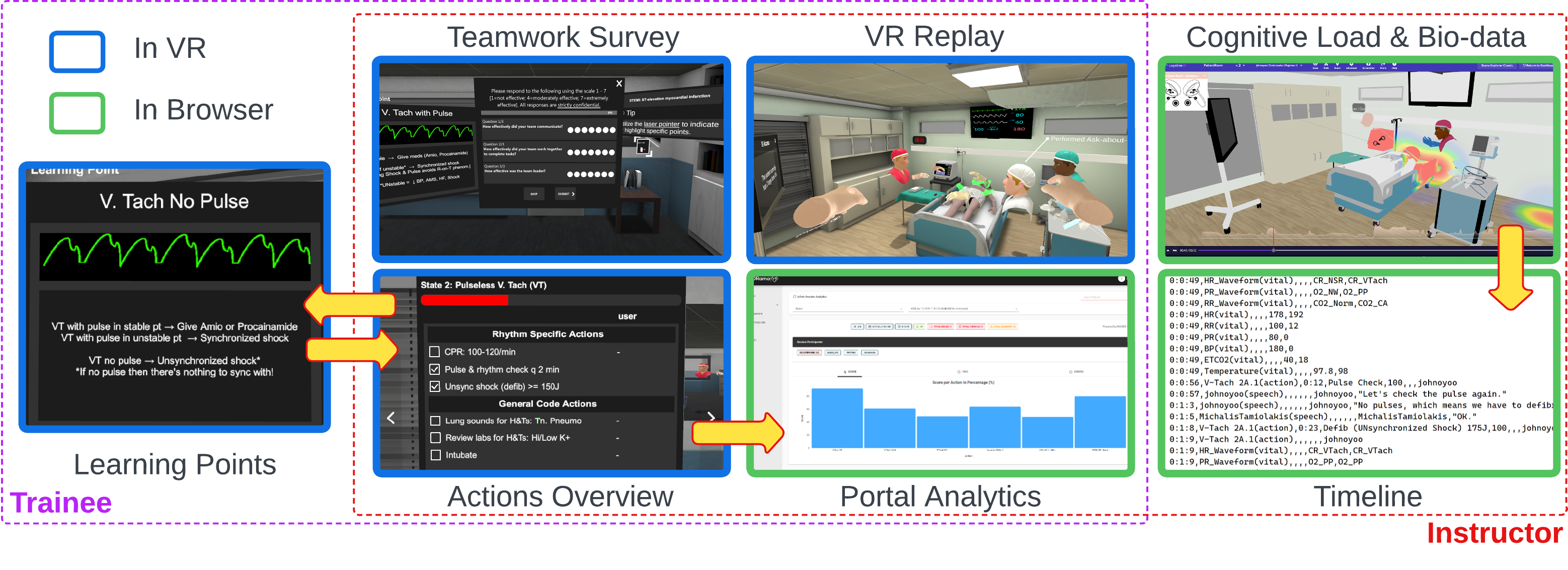}
    \caption{
    Our system analyzes collaborative VR medical training. This figure shows the post-session pipeline: (a) [purple] Trainees get performance insights via \textit{Actions Overview}, \textit{Teamwork Survey}, \textit{Learning Points}, \textit{VR Replay}, and \textit{Portal Analytics}.(b) [red] Instructors access the same data plus biometric metrics like \textit{cognitive load} and \textit{gaze}.
    }
    \Description{}
    \label{fig:teasCRM}
\end{figure*}

\textbf{CRM Assessment and Feedback:} iREACT provides both real-time and post-session feedback on CRM skills. Real-time feedback is provided through visual cues and auditory alerts for medication errors. Post-session feedback is delivered through the web Portal Analytics tool and in-VR debriefing tools (see Figure~\ref{fig:teasCRM}), providing a detailed timeline of events. This includes performed actions, state changes, vital signs, and transcribed speech logs, providing a comprehensive overview for thorough analysis. Specific CRM metrics include communication frequency, task distribution, response time and close loop communication. The \textit{Actions per Patient State} feature provides a visual summary of team performance for each stage of the scenario, indicating whether all required actions were completed. 
The \textit{Learning Points for Each State tool} offers concise feedback on correct actions and rationale to improve teamwork, communication, and decision-making skills.

To incorporate and visualize the tools described in this section, medical experts made crucial design considerations to ensure the user interfaces  display only critical information at specific moments. This approach prevents a potential overwhelming of the trainee with information and help trainees to maintain optimal performance under the high-pressure conditions.

\section{Formative Evaluation}
\label{sec:evaluation}

This research aimed to develop practical VR tools for understanding the complexities of team-based CA care by capturing richer data than traditional methods. iREACT records detailed user actions, communication, and cognitive load—metrics often inaccessible through conventional observation-based feedback.
While video-assisted debriefing can be helpful, it can be time-consuming to locate specific moments for review and may not capture all aspects of team performance. Moreover, features like the ability to record and replay VR sessions, along with the time-stamped communication log transcript, facilitate more comprehensive qualitative post-analysis. We posit that the tools presented have the potential to enhance CRM by equipping both team members and instructors with advanced resources to understand, assess, and improve team performance. 
As part of the initial testing for usability and educational affordances, we conducted a two-phase formative evaluation of iREACT: (1) a study with 12 clinicians, to understand the challenges and needs of ACLS instructors and trainees; and (2) a user case study with 5 trainees (distinct from the above experts) to examine the educational value of the new iREACT tools.

\textbf{Clinician Study:} For this study, we recruited a convenience sample of 12 experts comprised of: (i) seven ACLS instructors with experience leading CA teams (paramedics, nurses, emergency physicians, etc.) and (ii) five resident physicians with ACLS and CA experience. The study involved participation in the iREACT simulation and follow-up semi-structured interviews with each participant.  In the cognitive task analysis (CTA) portion of the study \cite{popov2025elucidating}, we elicited decision-making by having participants rewatch their performance from a first-person perspective and report on their perceived mental effort per each stage of cardiac arrest. The 250 unique cognitive processes identified in the CTA study using iReact simulation can be leveraged to develop training programs that accelerate expertise development (for more detail see \cite{popov2025elucidating}). In addition, they were asked questions about usability, realism, training effectiveness, potential challenges, and integration with existing training workflows. Instructors highlighted the need for real-time observation and feedback but noted limitations of current methods of video-assisted debriefing due to time constraints and difficulty navigating the video to quickly identify teachable moments. Detailed data on user actions and cognitive load was deemed essential for evaluating technical skills (e.g., CPR quality, medication selection) and critical thinking/stress levels, respectively. 
Evaluators proposed post-simulation feedback with action timeline markers, critical error summaries, and speech transcripts.

\textbf{User Case Study:} A total of 5 resident physicians  assigned to the hospital CA response team (called ``code team'') participated in an iREACT training session. Participants completed a post-training questionnaire assessing ease of use, perceived learning, engagement, and the effectiveness of the instructional/feedback tools integrated into iREACT. The questionnaire included both Likert-scale questions and open-ended questions to gather both quantitative and qualitative feedback. Due to space limitations, the questionnaires and a summary of responses from this user case study are provided as supplementary material. 

 Key results of this preliminary study are outlined in Table~\ref{tab:user_case_results}. Participants generally found the content to be moderately challenging and reported a moderate level of immersion. One participant commented: “Good immersive experience. I would like to try again with more time now that I'm more used to the technology.” The self-reported skill improvements after the session were notable, with the greatest gains in cardiac rhythm identification, clinical management, and communication. Participants valued the automated post-simulation feedback, which provided insights into their performance, error summaries, and key learning points relevant to their specific learning experience. 
Preliminary findings suggest an increase in how participants understand and use techniques to reduce cognitive load and manage situational
awareness. 
There were reports of several challenges during the session, including navigating the VR system, uncertainty about available resources, and uncertainty about the appropriate diagnosis and treatment. Despite this, most participants reported they would recommend iREACT for future CA code team members, suggesting it was an effective learning experience.

\begin{table}
    \centering
    \begin{tabular}{|c|c|}
        \hline
        \textbf{Topic} & \textbf{Result (avg)} \\
        \hline\hline
        Clinical challenge levels & 3.4 out of 5\\
        \hline
        Educational effectiveness & 3.6 out of 5\\
        \hline
        Level of immersion & 3.3 out of 5 \\
        \hline
        Unsure how to navigate in XR  & 5 out of 5 \\
        \hline
        Unsure of resources availability & 3 out of 5 \\
        \hline
        Unsure of proper diagnosis/treatment & 1/2 out of 5 \\
        \hline
         I would recommend to others & 4 out of 5\\
        \hline
    \end{tabular}
    \caption{Average key results from the User Case study conducted by 5 resident physicians.}
    \label{tab:user_case_results}
\end{table}

\section{Conclusion, Limitations \& Future Work}
\label{sec:Conclusion}

Overall, iREACT is a medical expert-designed toolset simulating the high-stress environment of a CA event. It provides personalized feedback on clinical decision-making, cognitive load, and attention, offering insights into teamwork, cognitive and behavioral factors, that influence clinical performance, and medical errors. iREACT is intended to complement traditional training methods.

While iREACT is a valuable VR training tool, it has certain limitations. First, it provides limited haptic feedback via controller vibrations, which may reduce realism for tactile skills such as CPR. However, this is unlikely to affect decision-making during Cardiac Arrest (CA) care—the system’s primary focus. Second, the current four-participant requirement may restrict flexibility. To mitigate this, we are investigating configurations with 2–3 participants assuming multiple roles, enhancing realism under low-staff conditions. Finally, initial user feedback noted minor challenges with VR interface navigation. However, this learning curve did not significantly impact training, as users adapted quickly and reported improved usability in subsequent sessions.

An extensive evaluation of iREACT is currently in progress at the University of Michigan Medical School. Future work includes the ability to adapt the iREACT platform and its analytic tools in other emergency care settings (e.g., Emergency Medical Services, trauma care, etc.). We are also exploring
AI methods to analyze the large, multi-source data sets generated by iREACT.

\balance 

\textbf{Acknowledgments.} 
This work was partially funded by the National Recovery and Resilience Plan "Greece 2.0" (TA$\Sigma\Phi$P-06378-REVIRES-Med), Innosuisse Swiss Accelerator (2155012933, OMEN-E), EU research and innovation program FIDAL (Horizon Europe GA No 101096146), the U.S. National Science Foundation (IIS-2202451), and the Doctors Company Foundation (AWD017372). We thank Achilleas Filippidis and John Petropoulos for their help at the very initial and early versions of this work.

\bibliographystyle{ACM-Reference-Format} 
\bibliography{bibliography}

\end{document}